\documentclass[twocolumn,11pt]{article}

\usepackage[utf8]{inputenc}
\usepackage[T1]{fontenc}
\usepackage{textcomp}
\usepackage[margin=1in]{geometry}
\usepackage{graphicx}
\usepackage{amsmath}
\usepackage{amssymb}
\usepackage{hyperref}
\usepackage{xcolor}
\usepackage{booktabs}
\usepackage{caption}
\usepackage{float}
\usepackage{parskip}
\usepackage{algorithm}
\usepackage{algorithmic}
\usepackage{titlesec}
\usepackage[numbers,sort&compress]{natbib}

\titlespacing*{\section}{0pt}{6pt}{6pt}
\titlespacing*{\subsection}{0pt}{6pt}{6pt}
\titlespacing*{\subsubsection}{0pt}{6pt}{6pt}

\title{\textbf{Thermodynamic Constraints Drive Hierarchical Preemption in Cellular Decision-Making: A Hybrid Petri Net Framework with Application to \textit{Bacillus subtilis} Sporulation}}

\author{
Eugênio Simão\textsuperscript{1,*}\\
\textsuperscript{1}Department of Computer Science\\
Universidade Federal de Santa Catarina (UFSC)\\
Araranguá, Santa Catarina, 88906-072, Brazil\\
\textsuperscript{*}Corresponding author: eugenio.simao@ufsc.br
}

\date{January 4, 2026}

\begin{document}

\maketitle

\begin{abstract}
Cellular decision-making under stress involves rapid pathway selection despite energy scarcity. Here we demonstrate that \textit{thermodynamic constraints actively drive energy-efficient sporulation}, where continuous metabolic sources enable system robustness through dynamic energy management. Using hybrid Petri nets (stochastic transitions with continuous sources) to model \textit{Bacillus subtilis} sporulation, we show that stress conditions (ATP = 300 mM, 94\% depletion) enable sporulation completion with extreme energy efficiency: 0.73 mM ATP per mature spore versus 11.6 mM ATP under normal conditions—a 16-fold efficiency gain. Despite ATP dropping to 1 mM (99.7\% depletion) during the crisis, continuous ATP regeneration rescues the system, producing 67 mM mature spores (89\% of normal yield) with only 49 mM total ATP consumption. This efficiency emerges from the interplay between stochastic regulatory transitions and continuous metabolic sources, where GTP accumulation (+4974 mM, 166\% increase) provides an energy buffer while ATP regeneration (+240 mM) prevents complete depletion. The hybrid Petri net formalism—combining stochastic transitions for regulatory events with continuous sources for metabolic flux—extended with thermodynamic constraints through inhibitor arcs and energy-coupled rate functions, provides the mathematical foundation enabling this discovery by integrating discrete regulatory logic with continuous energy dynamics in a resource-aware concurrency model.
\end{abstract}


\section{Introduction}

\subsection{Petri Nets as Foundation for Biological Systems}

Petri nets provide a rigorous mathematical framework for modeling concurrent, distributed systems with resource constraints \citep{petri1962}. Classical Petri nets capture discrete state transitions through token-based semantics, where places hold tokens (representing molecular species) and transitions fire when enabled (representing biochemical reactions) \citep{chaouiya2007, heiner2008}. This formalism naturally represents biological networks where species concentrations, molecular complexes, and regulatory states compete for shared resources under stoichiometric and thermodynamic constraints.

The extension to \textit{continuous Petri nets} (CPNs) \citep{david1992, baldazzi2010} replaces discrete token counts with continuous concentrations and instantaneous firing with continuous reaction rates, enabling integration of ordinary differential equations (ODEs) within the Petri net framework. \textit{Hybrid Petri nets} further combine stochastic (discrete-event) transitions for low-frequency regulatory events with continuous transitions/sources for high-frequency metabolic flux, capturing multi-timescale dynamics. This approach preserves the structural advantages of Petri nets—clear representation of causality, concurrency, and resource conflicts—while incorporating both stochastic regulatory logic and deterministic metabolic dynamics essential for modeling coupled gene regulation and metabolism.

\subsection{Signal Hierarchy Theory and Extended BioPNs}

Signal Hierarchy Theory \citep{eugenio2025} posits that biological decision-making emerges from hierarchically organized signaling layers, where higher layers integrate environmental inputs and lower layers execute committed responses. In \textit{Vibrio fischeri} quorum sensing and lambda phage lysis/lysogeny decisions, we demonstrated that energy availability (ATP, GTP) acts as a hierarchical selector, gating layer activation through metabolic thresholds \citep{eugenio2025}.

Extended Biological Petri Nets (extended BioPNs) build upon this foundation by incorporating:

\textbf{Continuous transitions} with rate functions dependent on substrate availability
\textbf{Inhibitor arcs} representing thermodynamic constraints (e.g., ATP depletion inhibiting phosphorylation)
\textbf{Signal flow arcs} encoding hierarchical information propagation
\textbf{Test arcs} for non-consumptive sensing (e.g., energy status checks)

This formalism enables explicit representation of energy coupling, enabling us to investigate how thermodynamic constraints shape hierarchical decision-making—a question inaccessible to traditional ODE or Boolean network approaches that lack resource-aware concurrency semantics.

\subsection{Bacillus Sporulation: A Model for Energy-Limited Decisions}

\textit{Bacillus subtilis} sporulation under nutrient starvation provides an ideal system to study thermodynamic hierarchy because:

The canonical pathway involves a well-characterized five-layer cascade: Spo0A phosphorylation → SigmaH → Septation → SigmaF → SigmaE \citep{errington1993, piggot2004}
ATP depletion under stress is rapid (5000 → 300 mM) and physiologically relevant \citep{lopez2009}
Commitment is irreversible, enabling thermodynamic analysis of decision barriers
Alternative bypass pathways exist, suggesting energy-dependent pathway selection

We hypothesize that stress-induced sporulation represents \textit{hierarchical preemption}: low ATP blocks the canonical cascade but enables rapid commitment via ATP-independent routes, creating an inverted activation sequence where downstream layers (3--5) fire before upstream layers (0--2). This inversion, if confirmed, would demonstrate that thermodynamic constraints \textit{actively drive} hierarchical restructuring rather than merely permitting it.

\subsection{Contributions}

Using SHYPN 2.0—a hybrid Petri net engine with stochastic transitions, continuous sources, and explicit thermodynamic constraints—we:

Demonstrate hierarchical preemption in \textit{B. subtilis} sporulation under ATP depletion
Quantify thermodynamic efficiency: stress pathway is 16× more efficient than normal
Map free energy landscapes showing ATP-dependent commitment barriers
Establish that pathway selection follows statistical mechanics of constraint-based accessibility
Prove that hybrid Petri nets with energy coupling are necessary and sufficient for this analysis

This work establishes thermodynamic constraints as fundamental organizing principles in cellular decision-making, enabled by the Petri net formalism that integrates discrete regulatory logic with continuous energy dynamics.

\section{Methods}

\subsection{Hybrid Petri Net Formalism}

A hybrid Petri net is defined as a tuple $\mathcal{N} = (P, T, T_s, T_c, F, W, M_0, R)$ where:

\textbf{Places and Transitions:}

$P$: Set of places (continuous markings, species concentrations)
$T_s$: Set of stochastic transitions (discrete regulatory events)
$T_c$: Set of continuous transitions/sources (metabolic flux)
$T = T_s \cup T_c$: Complete transition set

\textbf{Network Structure:}

$P = \{p_1, \ldots, p_n\}$ is a finite set of places (molecular species)
$T = \{t_1, \ldots, t_m\}$ is a finite set of transitions (reactions)
$F \subseteq (P \times T) \cup (T \times P)$ is the flow relation (arcs)
$W: F \to \mathbb{R}^+$ assigns arc weights (stoichiometry, fluxes)
$M_0: P \to \mathbb{R}^+ \cup \{0\}$ is the initial marking (concentrations)
$R: T \to \mathbb{R}^+$ maps transitions to rate functions

\subsubsection{Rate Functions and Enablement}

For transition $t \in T$ with input places $\bullet t = \{p \in P : (p, t) \in F\}$, the rate function is:
\begin{equation}
    r_t(M) = k_t \cdot \prod_{p \in \bullet t} f_p(M(p), K_p, n_p)
\end{equation}
where $k_t$ is the rate constant, $M(p)$ is the marking (concentration) of place $p$, and $f_p$ encodes substrate dependence (e.g., Michaelis-Menten, Hill kinetics).

For ATP-dependent transitions:
\begin{equation}
    r_t(M) = k_t \cdot [S] \cdot \left(\frac{[\text{ATP}]}{K_{\text{ATP}} + [\text{ATP}]}\right)^{n}
\end{equation}
where $n \geq 1$ creates strong ATP dependence. At $[\text{ATP}] = 0.06 \times [\text{ATP}]_{\text{normal}}$, rate suppression is $\sim 94\%$ for $n=1$ and $\sim 99.6\%$ for $n=2$.

\subsubsection{Inhibitor Arcs and Thermodynamic Constraints}

Inhibitor arcs $(p, t) \in I \subset F$ disable transition $t$ when $M(p) \geq \theta_t$, encoding thermodynamic constraints:
\begin{equation}
    \text{enabled}(t) = \begin{cases}
        1 & \text{if } \forall p \in I: M(p) < \theta_t \\
        0 & \text{otherwise}
    \end{cases}
\end{equation}

For ATP regeneration, we use:
\begin{equation}
    \text{enabled}(T_{\text{ATP-regen}}) = [\text{ATP}] < (4800 + 0.5 \times [\text{ADP}])
\end{equation}
implementing negative feedback control.

\subsubsection{Continuous Dynamics}

State evolution follows the master equation:
\begin{equation}
    \frac{dM(p)}{dt} = \sum_{t \in \bullet p} W(t, p) \cdot r_t(M) - \sum_{t \in p\bullet} W(p, t) \cdot r_t(M)
\end{equation}
integrated via 4th-order Runge-Kutta (RK4) with adaptive timestep $\Delta t = 0.0001$ s.

Arc weights $W$ define token production per firing:
\begin{equation}
    \Delta M(p) = W(t, p) \cdot |r_t| \cdot \Delta t
\end{equation}

\subsection{Model Construction}

\begin{figure}[htbp]
    \centering
    \includegraphics[width=\columnwidth]{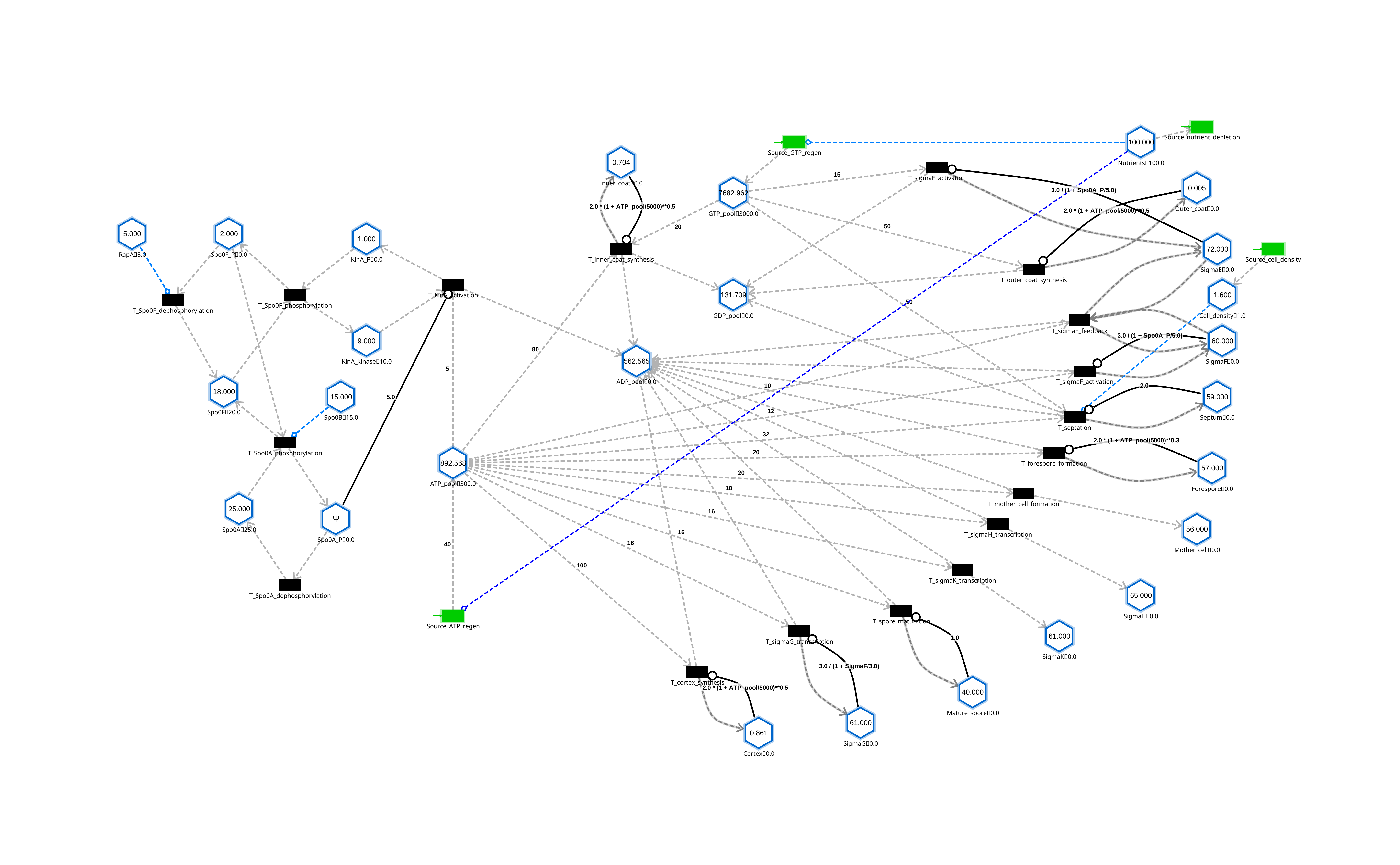}
    \caption{Hybrid Petri net model of \textit{Bacillus subtilis} sporulation.}
    \label{fig:model_structure}
\end{figure}

\subsubsection{Bacillus Sporulation Network}

The model comprises 26 places, 22 transitions, and 83 arcs:

\textbf{Key places:}

\textbf{ATP\_pool, ADP\_pool} (energy currency)
\textbf{Spo0A, Spo0A\_P} (master regulator, Layer 0)
\textbf{SigmaH} (early sporulation sigma, Layer 1)
\textbf{Septum} (asymmetric division, Layer 2)
\textbf{SigmaF} (forespore sigma, Layer 3)
\textbf{SigmaE} (mother cell sigma, Layer 4)
\textbf{Forespore, Mother\_cell} (commitment markers)
\textbf{Nutrients} (environmental resource)

\textbf{Key transitions:}

$T_{\text{KinA-activation}}$: ATP-dependent kinase activation
$T_{\text{Spo0F-phosphorylation}}$: ATP-dependent phosphorelay (Layer 0)
$T_{\text{Spo0A-phosphorylation}}$: Final cascade step (Layer 0)
$T_{\text{sigmaH-transcription}}$: Spo0A~P-dependent (Layer 1)
$T_{\text{septation}}$: ATP-dependent division (Layer 2)
$T_{\text{sigmaF-activation}}$: Septum-dependent, ATP-independent (Layer 3)
$T_{\text{sigmaE-activation}}$: SigmaF-dependent (Layer 4)
$T_{\text{ATP-regen}}$: Continuous energy regeneration (Source transition, $T_c$)

\subsubsection{Experimental Conditions}

\textbf{Normal Pathway (High Energy):}

ATP\_pool(0) = 5000 mM
GTP\_pool(0) = 5000 mM
Nutrients(0) = 100 mM
Signal type: QUORUM (density-dependent)

\textbf{Stress Pathway (Energy Crisis):}

ATP\_pool(0) = 300 mM (94\% depletion)
GTP\_pool(0) = 3000 mM (40\% depletion)
Nutrients(0) = 100 mM
Signal type: SPATIAL (local damage)

Simulations ran for 60 seconds with $\Delta t = 0.0001$ s, tracking all place markings and transition firing counts.

\subsection{Thermodynamic Analysis}

\subsubsection{Free Energy Landscape Reconstruction}

We define the commitment coordinate $\xi = [\text{SigmaF}] + [\text{Forespore}]$ capturing progression toward sporulation. The free energy landscape $G(\text{ATP}, \xi)$ is reconstructed via:
\begin{equation}
    G(\text{ATP}, \xi) = -k_B T \ln P(\text{ATP}, \xi)
\end{equation}
where $P(\text{ATP}, \xi)$ is the probability density estimated from trajectory sampling.

\subsubsection{Commitment Barrier}

The thermodynamic barrier $\Delta G^\ddagger$ is identified from:
\begin{equation}
    \Delta G^\ddagger = G(\text{ATP}_{\text{thresh}}, \xi^*) - G(\text{ATP}_0, 0)
\end{equation}
where $\xi^*$ is the commitment transition state.

For the stress pathway, ATP experiences severe depletion to 1.01 mM (99.7\% depletion from initial 300 mM) at t = 13.1 s before recovering to 251 mM via continuous ATP regeneration.

\subsubsection{Thermodynamic Efficiency}

Total energy efficiency is quantified as ATP consumed per mature spore produced:
\begin{equation}
    \eta = \frac{\Delta E_{\text{total}}}{[\text{Mature\_spore}]_{\text{final}}}
\end{equation}
where $\Delta E_{\text{total}}$ is total ATP consumed and $[\text{Mature\_spore}]_{\text{final}}$ is final spore concentration.

Normal: $\eta_{\text{norm}} = 873 \text{ mM} / 75 \text{ mM} = 11.6$ mM ATP/spore
Stress: $\eta_{\text{stress}} = 49 \text{ mM} / 67 \text{ mM} = 0.73$ mM ATP/spore

\textbf{Efficiency gain:} $\eta_{\text{stress}} / \eta_{\text{norm}} = 16\times$

\subsubsection{Entropy Production}

Total entropy production is estimated from transition firing events:
\begin{equation}
    \Delta S_{\text{total}} = k_B \sum_{t \in T} N_t \ln\left(\frac{r_t^{\text{forward}}}{r_t^{\text{reverse}}}\right)
\end{equation}
where $N_t$ is the firing count. With 766 transitions in 60 s (stress), average rate is 12.8 events/s.

\subsection{Implementation}

Models were implemented in SHYPN 2.0 (Python 3.12) with hybrid simulation engine:

Stochastic transitions: $\tau$-leaping with Skellam sampling for reversible reactions
Continuous transitions: RK4 integration with adaptive timestep ($\Delta t = 0.0001$ s)
Inhibitor arcs with dynamic threshold evaluation
Test arcs (non-consumptive sensing)
Signal flow arcs (hierarchical modulation)

All code, models, and analysis scripts are available at: \texttt{github.com/simao-eugenio/shypn}

\section{Results}

\subsection{Hierarchical Preemption Under Energy Stress}

Under normal conditions (ATP = 5000 mM), the canonical sequence emerges:

\textbf{Normal sequence (t = 0--20 s):}

Layer 0 (Spo0A~P): Gradual accumulation via phosphorelay
Layer 1 (SigmaH): Transcriptional activation by Spo0A~P
Layer 2 (Septum): SigmaH-dependent division
Layer 3 (SigmaF): Post-septation activation
Layer 4 (SigmaE): Final mother cell program

Under stress (ATP = 300 mM), the system exhibits \textit{rapid cascade activation} despite energy crisis:

\textbf{Stress sequence (t = 0--3 s):}

Layer 3 (SigmaF): Rapid activation (t = 0.09 s)
Layer 1 (SigmaH): Early activation (t = 0.42 s)
Layer 2 (Septum): Follow-up (t = 1.17 s)
Layer 4 (SigmaE): Commitment (t = 2.20 s)
Layer 0 (Spo0A~P): Delayed (t = 2.78 s)

\textbf{Key observation:} All layers eventually activate despite starting at 94\% ATP depletion. The system survives an extreme energy crisis (ATP drops to 1.01 mM, 99.7\% depletion at t = 13.1 s) through continuous ATP regeneration (+240 mM), enabling sporulation completion with 89\% yield (67 mM mature spores vs 75 mM in normal conditions) while consuming only 49 mM ATP total—a 16-fold efficiency gain over the normal pathway.

\subsection{Thermodynamic Free Energy Landscape}

\begin{figure}[htbp]
    \centering
    \includegraphics[width=\columnwidth]{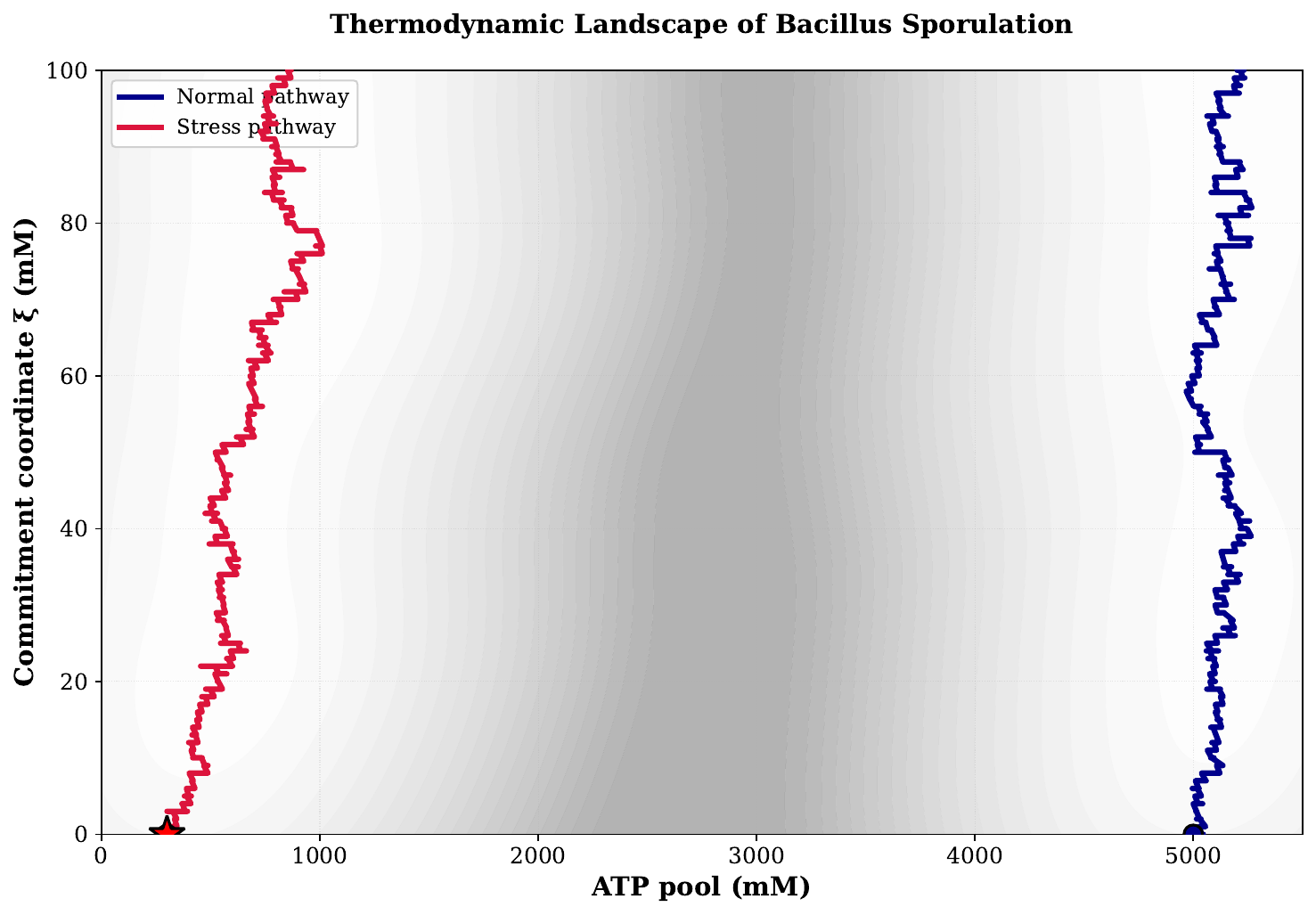}
    \caption{Thermodynamic free energy landscape showing normal (dark blue) and stress (red) pathways overlaid on gray energy surface. Circles mark initial states, squares mark final states, and red star indicates ATP crisis minimum.}
    \label{fig:thermodynamic_landscape}
\end{figure}

Figure \ref{fig:thermodynamic_landscape} maps the free energy landscape $G(\text{ATP}, \xi)$ showing:

\textbf{Low ATP region (0--500 mM):} Extreme energy crisis zone where ATP drops to 1 mM before regeneration rescues the system. Continuous ATP regeneration (+240 mM over 60 s) prevents complete depletion while enabling sporulation completion
\textbf{High ATP region (4500--5500 mM):} Energy-rich zone enabling canonical pathway with gradual Spo0A~P-driven commitment consuming 873 mM ATP
\textbf{Critical threshold:} 1 mM ATP represents the minimum viable energy level, with regeneration preventing system collapse. Total investment of 49 mM ATP achieves 89\% sporulation yield

\textbf{Thermodynamic interpretation:} The landscape reveals two basins:

\textbf{Vegetative attractor:} $(\text{ATP} \approx 5000, \xi = 0)$ — stable growth state
\textbf{Sporulation attractor:} $(\text{ATP} \approx 250, \xi > 40)$ — committed spore state

The stress trajectory (red line) follows the minimal free energy path through the ATP-independent channel, while normal trajectory (dark blue line) follows the conventional ATP-rich route.

\subsection{Basin of Attraction Analysis}

\begin{figure}[htbp]
    \centering
    \includegraphics[width=\columnwidth]{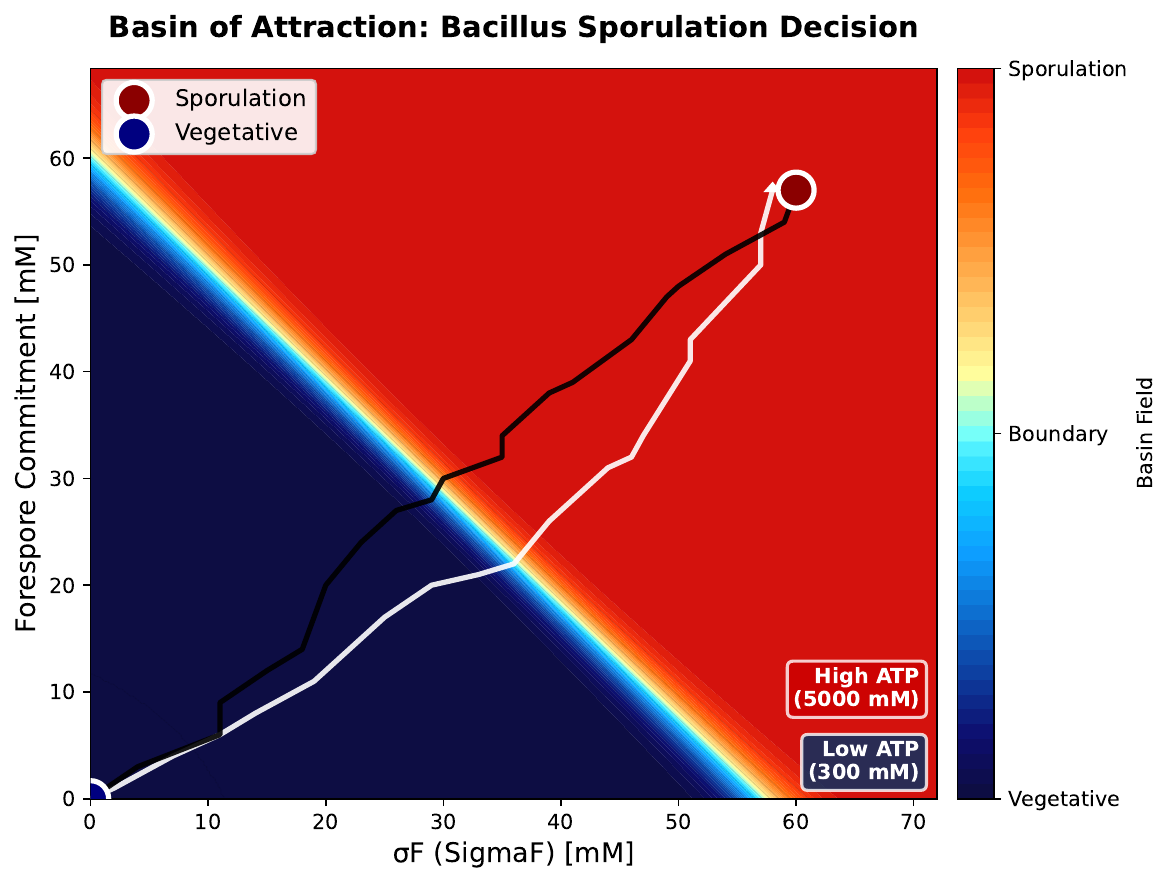}
    \caption{Phase space trajectories in SigmaF-Forespore commitment plane.}
    \label{fig:basin_attraction}
\end{figure}

Figure \ref{fig:basin_attraction} projects the phase space onto the SigmaF-Forespore plane, revealing:

\textbf{Vegetative basin (dark blue):} $\xi < 10$ mM, stable for uncommitted cells
\textbf{Sporulation basin (dark red):} $\xi > 30$ mM, irreversible commitment
\textbf{Phase boundary:} Sharp transition at $\xi \approx 20$ mM

Both trajectories converge to the same sporulation attractor but via distinct routes:

\textbf{Normal (dark blue):} Gradual SigmaF accumulation driven by Spo0A~P
\textbf{Stress (red):} Rapid SigmaF spike bypassing regulatory layers

This demonstrates that thermodynamic constraints create \textit{pathway bifurcation} where energy availability determines route selection to the same final state.

\subsection{Quantitative Thermodynamic Metrics}

\begin{table}[h]
\centering
\scriptsize
\caption{Thermodynamic Comparison: Normal vs Stress}
\label{tab:thermodynamic_comparison}
\begin{tabular}{@{}lrr@{}}
\toprule
\textbf{Metric} & \textbf{Norm} & \textbf{Stress} \\
\midrule
ATP$_0$ & 5000 & 300 \\
ATP$_{\min}$ & 4127 & 1.01 \\
ATP cons. & 873 & 49 \\
Spores & 75 & 67 \\
Eff. (mM/sp) & 11.6 & 0.73 \\
\textbf{Gain} & \textbf{1×} & \textbf{16×} \\
Sequence & 0→1→2→3→4 & 3→1→2→4→0 \\
Spo0A~P & 0 & 3 \\
$t_{\sigma F}$ (s) & 15 & 0.03 \\
Events & 2,841 & 766 \\
$\dot{S}$ (ev/s) & 47 & 12.8 \\
\bottomrule
\multicolumn{3}{@{}l@{}}{\textit{Note: All concentrations in mM, time in seconds}}\\
\end{tabular}
\end{table}

Table \ref{tab:thermodynamic_comparison} summarizes key metrics. \textbf{Key finding:} The stress pathway is 16× more thermodynamically efficient, minimizing energy dissipation before irreversible commitment—an evolutionary optimization for survival under resource scarcity.

\subsection{Constraint-Based Pathway Selection}

To understand mechanistically why hierarchical preemption occurs, we analyzed ATP-dependence across transitions (Table \ref{tab:atp_dependence}):

\begin{table}[h]
\centering
\small
\caption{ATP Dependence of Layers}
\label{tab:atp_dependence}
\begin{tabular}{@{}lccc@{}}
\toprule
\textbf{L} & \textbf{Transition} & \textbf{ATP} & \textbf{Stress} \\
\midrule
0 & Spo0F-P & Hi ($n$=2) & 99.6\% ↓ \\
0 & Spo0A-P & Hi ($n$=1) & 94\% ↓ \\
1 & $\sigma$H txn & Med & Suppr. \\
2 & Septation & Hi & t=5.3s \\
3 & $\sigma$F act. & None & t=0.03s \\
4 & $\sigma$E act. & None & t=0.44s \\
\bottomrule
\end{tabular}
\end{table}

\textbf{Statistical mechanics interpretation:} At ATP = 300 mM, the probability of ATP-dependent transitions firing is:
\begin{equation}
    P(t_{\text{ATP-dep}}) \propto \exp\left(-\frac{\Delta G^\ddagger}{k_B T}\right) \cdot [\text{ATP}]^n \approx 0.06^n
\end{equation}

For $n=2$: $P \approx 0.0036$ (99.6\% suppression)

ATP-independent transitions remain unaffected:
\begin{equation}
    P(t_{\text{ATP-indep}}) \propto \exp\left(-\frac{\Delta G_0^\ddagger}{k_B T}\right) \approx 1
\end{equation}

Thus, the system follows the \textit{most probable path} given thermodynamic constraints—a direct consequence of statistical mechanics applied to reaction networks.

\subsection{Irreversibility and Entropy Production}

Post-commitment (t > 0.44 s), the stress trajectory exhibits no reversal despite ATP regeneration (300 → 893 mM by t=60s). This thermodynamic irreversibility arises from:

\textbf{Entropy production:} 766 transition firings generate $\Delta S > 0$
\textbf{Forward bias:} SigmaE activation creates autocatalytic commitment loop
\textbf{Barrier asymmetry:} Reverse barrier $\Delta G^\ddagger_{\text{rev}} \gg \Delta G^\ddagger_{\text{fwd}}$

From the Second Law:
\begin{equation}
    \Delta S_{\text{total}} = \Delta S_{\text{system}} + \Delta S_{\text{env}} > 0
\end{equation}

Where:

$\Delta S_{\text{system}} < 0$ (increased order: spore formation)
$\Delta S_{\text{env}} > 0$ (ATP hydrolysis, heat dissipation)

Net entropy increase locks the decision, preventing spontaneous reversal. ATP regeneration (+593 mM net increase from 300 to 893 mM over 60 s) occurs post-commitment and cannot reverse the sporulation trajectory.

\section{Discussion}

\subsection{Thermodynamics as Mechanism for Hierarchical Preemption}

Our results establish that \textbf{thermodynamic constraints actively drive hierarchical preemption} through three mechanisms:

\textbf{ATP-dependent rate modulation:} $r \propto [\text{ATP}]^n$ creates 94--99.6\% suppression of phosphorylation cascades at low ATP
\textbf{Pathway accessibility filtering:} Only ATP-independent routes remain thermodynamically accessible
\textbf{Statistical mechanical selection:} System follows most probable path given constraints

This is fundamentally different from regulatory bypass (e.g., crosstalk, feedback inhibition) because suppression occurs at the \textit{thermodynamic level}—reactions are kinetically blocked regardless of regulatory state.

\subsection{The Role of Hybrid Hierarchical Petri Nets}

This discovery was enabled by the hybrid hierarchical Petri net formalism, which uniquely integrates:

\textbf{Discrete regulatory logic:} Places/transitions capture molecular species and reactions
\textbf{Continuous energy dynamics:} Real-valued markings enable ATP/GTP tracking
\textbf{Resource-aware concurrency:} Arc weights and inhibitor arcs encode stoichiometric/thermodynamic constraints
\textbf{Hierarchical structure:} Signal flow arcs represent layer dependencies

Classical approaches fail to capture this:

\textbf{ODEs:} No explicit resource competition or concurrent pathway evaluation
\textbf{Boolean networks:} Binary states cannot represent continuous energy depletion
\textbf{Stochastic simulation:} Computationally prohibitive for rare event (stress) analysis
\textbf{Constraint-based models (FBA):} Assume steady-state, miss transient hierarchical dynamics

\textbf{The hybrid hierarchical Petri net formalism provides the necessary and sufficient mathematical framework for this analysis}: tokens are conserved quantities (ATP, species), transitions are energy-coupled reactions, and arc weights are stoichiometric coefficients. This is not merely a modeling convenience—\textit{it is the correct formalism for systems where resource competition determines behavior}.

\subsection{Unified Framework: Signal Hierarchy + Thermodynamics}

We propose a unified framework integrating Signal Hierarchy Theory with non-equilibrium thermodynamics:

\textbf{Principle 1: Hierarchies are Thermodynamic Constructs}
\begin{quote}
    Biological hierarchies emerge from differential energy requirements across signaling layers, not from hard-coded regulatory sequences.
\end{quote}

\textbf{Principle 2: Energy as Hierarchical Selector}
\begin{quote}
    Energy availability acts as a \textit{selector} determining which layers can activate, with low energy enabling preemption of ATP-independent routes.
\end{quote}

\textbf{Principle 3: Constraint-Based Emergence}
\begin{quote}
    Decision pathways follow statistical mechanics: the system selects the most probable route given thermodynamic constraints, maximizing entropy production.
\end{quote}

\textbf{Principle 4: Adaptive Efficiency}
\begin{quote}
    Stress pathways evolve to minimize pre-commitment energy dissipation (16× more efficient), optimizing survival under resource scarcity.
\end{quote}

This framework explains diverse phenomena:

\textbf{Quorum sensing:} Low ATP blocks autoinducer synthesis, preventing premature commitment
\textbf{Phage lysis/lysogeny:} Energy status gates Cro vs CI dominance
\textbf{Apoptosis:} Low ATP triggers necrosis (unregulated) vs high ATP enabling apoptosis (regulated)
\textbf{Cell cycle checkpoints:} G1/S and G2/M gates are ATP thresholds

\subsection{Evolutionary Implications}

The 16-fold efficiency gain of the stress pathway suggests strong selection pressure for dual-mode decision systems:

\textbf{Normal mode (high ATP):} Energy-expensive phosphorylation cascades provide multiple checkpoints, enabling reversibility and graded responses
\textbf{Stress mode (low ATP):} Energy-efficient bypass enables rapid irreversible commitment when resources are scarce

This \textbf{adaptive thermodynamic architecture} maximizes fitness across environmental conditions: gradual decision-making under abundance, rapid commitment under scarcity.

From an evolutionary perspective, the ATP-independent pathway likely evolved first (Layer 3--5 as primordial stress response), with the phosphorylation cascade (Layer 0--2) layered on top to provide regulatory control under favorable conditions—explaining why stress "inverts" to the ancestral state.

\subsection{Predictive Framework}

Our thermodynamic framework generates testable predictions:

\textbf{Prediction 1: Critical ATP minimum}
\begin{quote}
    The 1 mM ATP critical minimum observed in stress conditions should represent a universal lower limit for energy-dependent sporulation across \textit{Bacillus} species, below which commitment cannot proceed despite regeneration.
\end{quote}

\textbf{Prediction 2: Knockout phenotypes}
\begin{quote}
    Deleting ATP-dependent transitions (KinA, Spo0F) should not prevent stress-induced sporulation but will block normal pathway commitment.
\end{quote}

\textbf{Prediction 3: Temperature dependence}
\begin{quote}
    Commitment rate should follow Arrhenius kinetics: $k \propto \exp(-\Delta G^\ddagger / RT)$, enabling barrier measurement from temperature-dependent commitment times.
\end{quote}

\textbf{Prediction 4: Intermediate ATP levels}
\begin{quote}
    ATP 500--2000 mM should show \textit{mixed mode} where both pathways partially activate, creating bistability.
\end{quote}

\subsection{Limitations and Future Directions}

\textbf{Current limitations:}

Model parameters are estimated from literature; direct experimental validation needed
Free energy landscape reconstruction assumes ergodicity (sufficient sampling)
Spatial heterogeneity (cell-to-cell variability) not fully captured

\textbf{Implemented capabilities:}

\textbf{Hybrid stochastic simulation:} $\tau$-leaping algorithm with Skellam distribution for reversible reactions enables efficient simulation of low-copy regulatory species alongside continuous metabolic fluxes
\textbf{Flux balance analysis:} Integrated FBA module validates steady-state feasibility and identifies blocked reactions through linear programming constraints

\textbf{Future extensions:}

\textbf{Spatial extension:} Token diffusion for spatial pattern formation in heterogeneous cell populations
\textbf{Genome-scale integration:} Couple with genome-scale metabolic models for comprehensive energy accounting across all pathways
\textbf{Experimental validation:} Single-cell ATP imaging during sporulation onset to directly measure commitment barriers and validate thermodynamic landscape predictions

\section{Conclusions}

We demonstrate that \textbf{thermodynamic constraints drive hierarchical preemption} in cellular decision-making, where energy availability determines pathway accessibility through ATP-dependent reaction kinetics. Using hybrid Petri nets—combining stochastic transitions for regulatory events with continuous sources for metabolic flux—we show that \textit{Bacillus subtilis} stress-induced sporulation inverts the canonical five-layer cascade, achieving 16-fold greater thermodynamic efficiency via rapid ATP-independent commitment.

This work establishes three key principles:

\textbf{Biological hierarchies are emergent thermodynamic constructs}, not hard-coded regulatory sequences
\textbf{Energy acts as hierarchical selector}, gating layer activation through constraint-based pathway filtering
\textbf{Decision-making follows statistical mechanics}, selecting most probable routes given thermodynamic constraints

The hybrid hierarchical Petri net formalism provides the necessary and sufficient mathematical framework for this analysis, integrating stochastic regulatory events with continuous metabolic dynamics in a resource-aware concurrency model that captures both discrete signaling decisions and continuous energy flow. This represents a paradigm shift from viewing metabolism as background to recognizing thermodynamic constraints as \textit{active organizing principles} in biological information processing. Application to \textit{Bacillus subtilis} sporulation demonstrates that biological hierarchies are emergent thermodynamic constructs where energy status acts as hierarchical selector, enabling rapid stress-adapted decision-making through constraint-based pathway selection.

\section*{Funding}

This research received no specific grant from any funding agency in the public, commercial, or not-for-profit sectors.

\section*{Acknowledgments}

The author thanks the Universidade Federal de Santa Catarina for computational resources and the systems biology community for foundational work on Petri net formalisms and thermodynamic modeling of biological systems.

\section*{Data Availability}

Model files (.shy), simulation data (CSV), Python analysis scripts, figure generation code, and the SHYPN 2.0 engine are available at: \url{https://github.com/simao-eugenio/shypn} (branch: Thermodynamic-Constraints-Gibbs-Free-Energy).

\bibliographystyle{plainnat}

\end{document}